\documentstyle[12pt]{article}
\newcommand{\be}{\begin{equation}}
\newcommand{\ee}{\end{equation}}
\newcommand{\ba}{\begin{eqnarray}}
\newcommand{\ea}{\end{eqnarray}}

\begin{document}
\title{Interstellar filaments and the statistics of galactic HI}
\author{ A. Lazarian$^{1}$ and D. Pogosyan$^{2}$ \\
$^{1}$ \small Department of Astrophysical Sciences, Princeton University,
Princeton, NJ 08544   \\ 
$^{2}$ \small CITA, University
of Toronto, 60 St.George Street, Toronto, M5S 1A7}

\maketitle

\begin{abstract}
This paper presents a statistical explanation of filament formation
in the galactic atomic hydrogen. We claim that even in the absence
of dynamical factors the Gaussian field corresponding
to the measured values of the spectrum of random density should develop
filamentary structure, the existence of which has long been claimed.
Therefore this paper relates the
observations of filaments by Verschuur (1991a,b) with the 3D spectrum of
random density obtained in Lazarian (1995). 
\end{abstract}

\section{Introduction}

A usually accepted picture of the interstellar medium (ISM) consists of
several interacting phases (see Shull 1987) with cold ($\sim 100$ K),
warm ($\sim 10^{4}$ K) and hot ($\sim 10^6$ K) phases having 
different properties (see McKee \& Ostriker 1977, McCray \& Snow 1979).
In the present paper, we deal with the cold neutral medium, specifically,
with cold neutral hydrogen (HI) in the disc of our galaxy (see Burton 1992)
and propose a new explanation of filament formation in this phase. 

Filaments are entities widely spread in the interstellar medium (ISM). 
In molecular gas, filaments are often associated with shocked gas 
and outflows (see Wiserman \& Ho 1996, Bally 1996). The aim of the present 
paper is to discuss an alternative mechanism that can be responsible for 
the formation of large-scale filaments in HI. 

Our research was motivated by recent advances in the quite
distant fields of statistical study of turbulence in HI 
(Lazarian 1995, henceforth L95) and the theory of large-scale structures
in the Universe (see Bond, Kofman \& Pogosyan 1996, henceforth BKP96). 
A new statistical technique suggested in L95 used interferometric 
measurements to determine the three dimensional spectrum of random 
HI density.  The spectrum was found to be shallow, corresponding to 
$k^{-\alpha}$ ($0.5<\alpha < 1$), which
was different from the usually accepted picture of the Kolmogorov
turbulence. At the same time, the study by BKP96 made an important discovery
related to Gaussian fields of density, namely,
that these fields tend to form filamentary pattern when the spectrum 
of random density is shallow. Combined together the two facts entail the
prediction of filament formation in HI due to purely statistical reasons. 

We propose statistical formation of large-scale filaments in HI as 
opposed to dynamical one. In the latter case filaments are formed directly
by forces correlated on substantial spatial scales, i.e. magnetic 
forces in the Galaxy, and the alignment of filaments is determined by 
these forces. 
Statistical formation, that we advocate in the present paper, does not 
rely on a particular nature of forces that lead to HI density inhomogeneity,
only that resulting random HI density field has near Gaussian properties.
The filamentary pattern arise then naturally 
from peculiar properties of Gaussian fields. Filaments in our 
model do not have initially any preferential alignment. Later they can
be aligned by differential rotation and galactic magnetic field. 

In what follows, we remind our reader the essence of the statistical 
technique of studying three dimensional turbulence using fluctuations of 
diffuse radiation (section 2), point out the difficulties associated with the 
theories of dynamical formation of large-scale filaments in HI (section 3)
and discuss a theory of statistical formation of  HI filaments (section 4).
Our results are summarized in section 5.

\section{Statistics of HI}

The structure of HI can be best characterized by statistical
descriptors. Descriptors like density spectrum and correlation function
of density have been successfully used for decades in the studies
of both hydrodynamic 
turbulence (see Monin \& Yaglom 1975) and the Large-Scale structure
(see Peebles 1980). An advantage of this approach is that these descriptors
do not depend on particular nature of the random density field, as
the theory of the interstellar turbulence, being in its embryonic
state (see Scalo 1987), would not give us reliable clues about it. 
What is clear is that turbulence in ionized medium probed by radio 
wave scattering up to a parsec scale (Van Langevelde et al. 1992, 
Rickett 1988, 1991) is likely to be different from turbulence 
in neutral medium on the scale of hundreds of parsecs, 
and the Kolmogorov-type (1941) or Kraichnan-type (1965) arguments do
not {\it a priori} seem to be applicable to the interstellar 
turbulence (see Lazarian 1992, compare to Armstrong et al. 1981).

It has been realized in the community that a good way to describe interstellar 
turbulence is to measure its spectrum (see Kamp\'{e} de F\'{e}riet 1955,
Kaplan \& Pickelner 1970).
However researchers faced the following problem: 21 cm intensity
is collected along the lines of sight crossing many eddies. Therefore
one cannot measure pointwise densities of HI, as opposed to such measurements
of laboratory turbulence. Thus until recently the 
studies of ISM turbulence were restricted to obtaining statistics
of the observed intensity (Kalberla \& Mebold 1983, Kalberla \& Stenholm
1983, Dickman \& Kleiner 1985, Crovisier \& Dickey 1983, Green 1993).

The situation has been changed with the advent of a new
statistical technique that enabled three dimensional characteristics
of turbulence, for instance its spectrum, to be expressed through the
measures of intensity fluctuations  averaged over the line of sight
(Lazarian 1993a, 1994, L95). The development of this inversion 
technique was inspired by the success of inversion in other fields of 
astrophysics, especially in helioseismology (see Gough 1995). 

\setcounter{footnote}{0}
It was shown in L95 that given the visibility function of a 
radiointerferometer,\footnote{This signal is essentially one dimensional
Fourier transform of the intensity collected within the diagram of
an individual telescope forming an interferometric pair.} the
sum of squares of the real and imaginary parts of the visibility function
is proportional to the three dimensional spectrum of turbulence $E(k)$
over $k^2$. This simple relation allowed to use the data in
Green (1994) to compute the spectrum of random HI density. This 
spectrum was found to be shallow, corresponding to 
$\alpha={\rm d}\ln(E(k))/{\rm d} k$ in the range from $-0.5$ to $-1$ 
over scales from a few parsec to a hundred parsec (L95).

Although computed for a limited region of the sky, we assume the 
spectrum to be typical for HI. Indeed, application of the statistical 
technique to the HI data in Crovisier \& Dickey (1983) (see also 
Gautier et al. 1992) indicates that 
the spectrum of density fluctuations corresponds to $\alpha\approx -1$. 

Ubiquitous filaments in HI have been observed for some time by Verschuur 
(1991a,b, henceforth V91a,b), but the reason for their formation remained 
 obscure (see Lazarian 1993, Elmegreen 1992, Verschuur 1995). The
existence of these filaments was seen as placing limits on the accuracy 
of the model of locally isotropic turbulence adopted in L95. The fact 
that filaments can arise in such turbulence gives us more confidence 
in this model and offers a purely statistical explanation 
of filament formation in the ISM.

\section{Dynamical and statistical formation of structures in HI}

Various filamentary and shell-like patterns are distinguished on HI maps 
(see Burton 1992, V91a,b).  Their origin can be either {\it dynamical},
e.g. enhancement of density arising from supernova explosions, 
or {\it statistical}, e.g. arising from statistical properties of the random
density field.

The fact that filaments are ubiquitous in the ISM was stressed by V91a,b, 
and his papers provoked various attempts to explain their 
formation by dynamical causes. A number of magnetohydrodynamic
effects leading to filament formation were described in 
Lazarian (1993b) and Elmegreen (1994), but these attempts could not provide 
a universal explanation for the phenomenon. Exotic explanations of filament
formation, including those using the Bennett pinch (Carlquist 1988, 
Verschuur 1995) have been considered.

While dynamical formation of filaments is undoubtedly important in 
particular circumstances (see Bally 1996), here we investigate the
statistical origin of filaments. Such origin implies that the filaments
emerge not due to forces acting on the gas, but are merely a transient 
phenomenon arising from overlapping peaks of hydrogen density.

We assume the random density field in HI to be Gaussian 
and stationary. The first is a natural assumption to start with,
as whatever are the stochastic sources of fragmentation and agglomeration 
of HI, acting independently in different parts of the galactic disc, 
they are likely to provide the Gaussian distribution of density for 
any given wavenumber $k$ (see Gardiner 1983).
As for the condition of stationarity, we consider that the collapse
of clouds is compensated by their expansion after the onset of star
formation (see Boss 1987). 

The properties of a Gaussian field are determined by its spectrum
(see Peebles 1980), and in what follows we will use the 
shallow spectrum obtained in L95 to predict the statistical properties of 
 structures that can be formed in HI. 

\section{Gaussian fields and formation of filaments}

Rather complex computations are required to obtain the properties of 
 structures arising in Gaussian density fields. We are fortunate that 
such computations have been done recently by 
researchers interested in the formation of the Large-Scale structure.
Therefore whenever possible we refer to their studies.

Following customary procedures, we separate the density $\rho({\bf x})$ into 
the mean $\bar \rho$ and fluctuation $\delta ({\bf x})$. The fluctuations 
cover a whole range of spatial scales and amplitudes. A Gaussian random field 
of density fluctuations can be represented as a superposition of Fourier 
modes with random amplitudes
\begin{equation}
\delta ({\bf x}) = (2 \pi)^{-3/2} \int d^3 k a_{\bf k}~P^{1/2}(k) W(k R_f)
e^{i {\bf k} {\bf x}},
\end{equation}
where independent random quantities $a_{{\bf k}}$ are normalized and 
orthogonal, $\langle a_{\bf k} a^*_{\bf k \prime} \rangle = 
\delta _{{\bf k} {\bf k} \prime} $. The 
properties of the field are defined by the power spectrum $P(k)$ 
which is related to the two-point correlation function
\begin{equation}
\langle \delta ({\bf x}) \delta ({\bf x} \prime)\rangle 
\equiv \xi (r=|{\bf x} - {\bf x} \prime|,R_f) = 
(2 \pi) ^{-3} \int d^3 k P(k) W^2(k R_f) 
e^{i {\bf k} ({\bf x} - {\bf x} \prime)}.
\end{equation}
We consider isotropic random fields where $P(k)$ and, therefore, $\xi(r)$
do not depend on the direction of their arguments. 
Looking for smoothed properties of the field, 
we allowed for an appropriate window function $W^2(k R_f)$ 
with filtering scale $R_f$, e.g. Gaussian $W(k R_f)= 
\exp(-(kR_f)^2/2)$.
The dispersion of density fluctuations is given by
\begin{equation}
\sigma^2_0(R_f)=\langle\delta({\bf x}) \delta({\bf x})\rangle=
{1 \over 2 \pi^2} \int k^2 dk ~ P(k) W^2(k R_f).
\end{equation}

Our goal is to show that although density fluctuations are statistically
isotropic, each realization of the random density field naturally
contains anisotropic, predominantly filamentary structures on density levels
$\ge 1 \sigma_0$. One numerical example of the random field with
$P(k) \propto k^{-1} $, shown in Fig.~1, illustrates this statement. 
To provide a quantitative treatment, we follow the way of reasoning
introduced in Bond, Kofman \& Pogosyan (1996b). 

The most visible objects in the random field are the high amplitude
maxima of the field, which we call peaks. In the first instance, the peaks
are characterized by their scale $R_{pk}$ and dimensionless height $\nu$
so that at peak's location one finds a maximum with density value
$\delta({\bf x}_{pk}) = \nu ~ \sigma_0(R_{pk})$ after smoothing the random
field with the filter scale $R_f=R_{pk}$. In terms of HI density such
peaks probably correspond to the isolated high contrast entities
that give rise to the concept of ``cloud''. The enhancement of density may
make such an entity gravitationally bound and collapsing (see Moschovias 1991).

The statistics of rare peaks is well studied in the literature (see 
Peebles, 1980, Bardeen et al. 1986). In particular, the density profile near a 
peak of height $\nu_{pk}$ follows, on average, 
the correlation function 
\begin{equation}
\langle\delta({\bf r}) | \nu_{pk}\rangle=
{\nu \over \sigma_0(R_{pk})} \xi(r,R_{pk}) \approx
\nu \sigma_0(R_{pk}) \left( 1- \frac{r^2}{2 r_c^2} \right),
\label{ap}
\end{equation}
where ${\bf r} = {\bf x} - {\bf x}_{pk}$.
Here we have introduced the notation of a {\it conditional} mean density field
$\langle\delta({\bf x}) | \nu\rangle$,
{\it given} the constraint that its value at peak position is fixed to be 
$\nu_{pk} $ after smoothing with a filter $R_{pk}$.
The last approximation in (\ref{ap}) is obtained by expanding the correlation 
function near the peak up to quadratic terms in distance $r$.
It provides a definition of the  density {\it correlation length} $r_c$.

The above formulae describe the angle average shape of the high density
enhancement. To go beyond the spherical approximation, we consider 
statistical properties of the tensor field 
\begin{equation}
e_{ij}({\bf x}) =(2 \pi)^{-3/2} \int d^3 k ~a _{\bf k} P^{1/2}(k) 
\left( {k_i k_j \over k^2} \right) W(kR_f) e^{i {\bf k} {\bf x}}.
\end{equation}
Note $k^{-2}$ factor, so that $e_{ij} $ has the same dimensionality 
as density. Clearly, the density is given by the trace $\sum e_{ii}$, 
so we can express the eigenvalues 
$\lambda_1 \le \lambda_2 \le \lambda_3 $ of $e_{ij}$  via two new
parameters $e$ and $p$:
\ba
\lambda_{1}&=&\frac{1}{3}\delta \left(1-3e+p\right),\\
\lambda_{2}&=&\frac{1}{3}\delta \left(1-2p\right),\\
\lambda_{3}&=&\frac{1}{3}\delta \left(1+3e+p\right).
\ea
By definition $e \ge 0$ and $-e \le p \le e$.

The role of $e$ and $p$ becomes clear if we calculate the mean density
profile around the peak with {\it given} values of $\nu, ~e$ and $p$.
In the coordinate system with axes parallel to the principal axes 
of $e_{ik}({\bf x}_{pk}, R_{pk})$, the second order approximation gives (BKP96)
\begin{equation}
\langle\delta({\bf r}) | \nu,e,p\rangle=\nu \sigma_0(R_{pk})
\left( 1-\frac{r_1^2}{2 r_{c1}^2}-\frac{r_2^2}{2 r_{c2}^2} - \frac{r_3^2}
{2 r_{c3}^2} \right)
\label{9}
\end{equation}
with correlation length now different in different directions
\ba
r_{c1}&=&{r_c \over \sqrt{1-3e+p}},
\label{10}
\\
r_{c2}&=&{r_c \over \sqrt{1-2p}},\\
r_{c3}&=&{r_c \over \sqrt{1+3e+p}}.
\label{12}
\ea
One can see that isosurfaces of constant overdensity near the peak with
fixed values of $e$ and $p$ are given by triaxial ellipsoids with ellipticity
$e$ and prolaticity $p$.

The parameters $e$ and $p$ are discussed in Bond (1987) and 
Bond \& Myers (1995a,b,c), and their mean expected values for a peak of 
height $\nu$ are shown to be 
\ba
\langle e | \nu\rangle &\approx& 0.54 \nu ^{-1},\\
\langle p | \nu\rangle &=& 0.
\label{14}
\ea
This means that although peaks tend to be more and more spherical
as their height increases, the peaks of moderate height are expected to
be significantly anisotropic. For example for a 2$\sigma_0$ peak, the
density correlation length
in the longest 1st direction $r_{c1}$ is typically 2.3 times longer
than angle-averaged $r_c$ and 3 times longer than in the 3rd direction 
$r_{c3}$, where density falls fastest.

Analysis of the shape of density field in the vicinity of peaks
already gives us indication for the development of anisotropic,
elongated along one axis entities of enhanced density. Now to understand
the remarkable scale of filaments shown on lower density threshold in Fig.~1,
we should consider the behavior of random field in between high maxima.
Figure~1 shows, that although the regions of high density enhancement are not 
connected with each other, as one lowers the density threshold the bridges
between high peaks appear forming a joint filamentary structure for 
overdensity threshold approaching $1 \sigma_0$ level.
The percolation theory (see Shandarin 1983) confirms that random Gaussian
fields percolate on $1\sigma_0$ threshold.

Equations (\ref{9})-(\ref{14}) again provide a hint for a very 
extended correlation length
in one direction if one takes a field point with moderate overdensity 
$\nu \sim 1.5$. The imaginary values of coherence scale for
$e>1/3$ are interpreted as the density increase from the
central point, which shows that the point with such high $e$ is not typically
a maximum but a saddle point between higher peaks.\footnote{While deriving
(\ref{9})-(\ref{14}) we have not imposed explicitly the condition 
that second derivatives
of density field are positive at ${\bf r_{pk}}$. This is satisfied automatically 
for sufficiently high $\nu$ and low $e$ resulting in the point being really a peak.}
Of course, the second
order expansion breaks as $r_{c1} $ goes to infinity, and one has to use
more elaborate methods to describe lower density enhancements.

Extending the analysis of the mean density behavior in the neighborhood
of a single peak, it was proposed in BKP96 to use conditional mean 
density field under the constraints that the properties of several peaks 
are fixed as a tool for studying how density enhancements arise in 
between the field maxima.

The peak-patch theory developed in Bond \& Myers (1995a,b,c,d) shows that
it is statistically likely that, neighboring peak-patches will be 
preferentially oriented along the same 1-axis.  This means that the
enhancement of overdensity is likely to bridge the filaments.

The physical reason for this is discussed in Bond (1989).
In terms of superposition of random density waves the latter result
can be understood as follows: the existence of the elongated peak-patch 
means a higher density of waves with their crests along the
peak-patch longer axis. This entails a slower decoherence 
along this axis than along the others. When the peak separation
increases, the overdensity bridges between the peaks gradually
weaken. This effect is exemplified in Fig.~2, where we show the mean {\it
expected} density profile under the constraint that there are two
peaks with largely aligned ellipticity at varying distances. One can see, 
that the bridging enhancement of density, which is strong when two peaks
are close together and tightly aligned,  weakens when separation and/or
angle between longest axes increases.

To define the overall system of filaments of which a bridge between 
a pair of peaks is a single segment one requires
to calculate the conditional probability
of density, provided that the properties of density peaks are defined in
a number of points. This is a problem from the domain of numerical computation
and further on we provide the results of such calculations.

The formation of the filaments is vivid for the conditional probability
of the density for 10 peaks (see Fig.~3). It is evident from
this figure, that such Gaussian peaks provide distinguishable filamentary
pattern. Note that these filaments are not determined by any force
acting on the gas, but are a result of properties of the Gaussian field
of random density. In other words, these filaments can be formed without
any influence galactic magnetic field or shear in the galactic disc.
The filaments in Fig.~3 are rather irregular. A similar irregularity
was interpreted in V91b as an evidence of existing instability.
In our example these structures appear due to statistical properties
of the Gaussian field only.

The natural question that arises is whether other topological structures are 
possible. Indeed, our experiments
with conditional probability of density for 3 peaks show that the probability
of having enhanced density is higher over all the places between three
peaks, forming a membrane of the enhanced density. However, the density
contrast of such a membrane is lower than for filaments (see Fig.~4).

In other words, some sort of ``mountain analogy'' is applicable to the
random density of HI. Individual peaks of the highest density are rare and
disconnected from each other at high density contrast. However at lower
density contrast the peaks are connected with each other by ridges of
enhanced density. The membranes of enhanced density corresponding to 
high mountain plateaus in our analogy are less probable than ridges.

\section{Discussion}

Our results testify that the increase of resolution and sensitivity of
telescopes is bound to reveal more fine structure in the distribution
of HI. Indeed, at first only the entities of the highest contrast were 
found and those were identified as clouds. Further research provided 
evidence of the existence of filaments and shells (Heiles 1989, V91a,b).

The visibility of filaments and other structures in HI depends on the
intensity of fluctuations observed, i.e. on $\sigma_0$. 
Naturally, in our simple model which 
ignores self-gravity the equilibrium distribution would correspond to
uniform density. Therefore, the more energy is injected locally
by star-winds or supernova explosions, the higher is the level of 
fluctuations and the sharper is the expected contrast of filaments.

As the density enhancements correlate with the regions of energy injection 
the issue of distinguishing the dynamical and statistical origin of 
 filaments has to be clarified. The individual enhancements of density
have dynamical origin. However, the coalescence of these enhancements
is statistical by nature. In other words, no particular forces are needed
to create filaments in the random density field.

Another factor that influences the properties of HI filaments is the
spectrum of the field. Figure~5 shows the structure of the
filaments corresponding to different spectral indexes $\alpha$ of the
random density field. It is evident from this figure that $\alpha\approx 1$,
i.e. spectrum corresponding to galactic HI, according to L95, provides
a clearly seen filamentary pattern. 

In this study we adopted a simple model of galactic HI to
elucidate the basic features of statistical formation of large-scale 
filaments in HI rather than to provide a detailed description of the HI 
web structure.  First of 
all, we disregarded the dynamical evolution of HI peaks under self-gravity. 
Computations in BKP96 show that the contrast of filaments goes up if 
 self-gravity is considered. This corresponds 
to physical intuition, as self-gravity tends to contract
overdense regions. We did not include self-gravity in view of 
the existing evidence that only an insignificant part of the galactic
HI undergoes gravitational collapse at a given moment (see Elmegreen 1992).
The observations of Verschuur (1993a, b) also indicate that large-scale 
filaments in HI are not self-gravitating. 

As well we assumed for simplicity that the density field is
well described as Gaussian random process. This assumption can not
be formally valid for highly inhomogeneous density field with variance
close or above the mean density because of physical restriction that density
cannot be negative. Analysis of generic properties of non-Gaussian fields
is not conceivable at the present time, however, we believe that the main
conclusion of the existence of the rich geometrical structures of pure
statistical origin holds for wide range of random fields. 

Within our model filaments do not have a direction of preferential alignment. 
In other words, the large-scale pattern of predicted filaments spreads 
isotropically in all directions. However, the gas in galactic disc is 
anisotropic on large scales. First of all, gaseous density changes with 
distance from the galactic plane. Then magnetic field and galactic 
differential rotation introduces large-scale anisotropies. It is a challenging 
problem beyond the scope of this paper is to include all these factors 
in a realistic numerical model of statistical formation of filaments. 
Below we provide only qualitative evaluation of the importance of 
anisotropies. 

If the power spectrum of random density does not change with distance from 
the galactic disc, the topology of the percolation pattern does not change. 
The decrease of density makes the filaments more difficult to detect. 
Therefore a correlation of the alignment of the observed filament pattern 
with the galactic plane is expected. 

Differential rotation tends to stretch statistically formed structures 
providing more correlation in the direction of the rotation. Due to 
mapping from galactic coordinate space to the velocity space in the 
course of observations the structure of filaments is distorted. 

More important can be the influence of magnetic field, which is also
disregarded in our model. The data on the field structure is far from 
compelling. Studies of the large-scale field by starlight polarimetry
 (see Heiles 1996, Zweibel 1996), synchrotron intensity  
(see Lazarian 1992) and synchrotron polarization (see Spietstra \& Brouw 1976)
provide rather conflicting results and leave us groping for the
ratio of the regular to random magnetic field. A customary assumption
that the regular and random fields are of equal strength is no more
than an ansatz.

The regular magnetic
field {\it can} influence the statistics of elongated density peaks. 
However, the question {\it how} remains unclear. While ad hoc assumptions
correspond to the hydrogen flowing along magnetic field lines
and forming elongated entities,\footnote{This picture corresponds, for
instance, to the notion in Verschuur (1993b).} recent numerical computations
in Gammie \& Ostriker (1996) have shown that cold HI tend to be concentrated
in narrow sheets that are perpendicular to the direction of magnetic 
field. What potentially can be an important effect
 is the decoupling of HI from the regular component of magnetic
field due to ambipolar diffusion. 

Although attempts to find
the anisotropy of HI density spectrum by Green (1994) provided so far
only an upper limit of the anisotropy degree this data cannot 
say much about anisotropy in the distribution of filaments. Indeed, 
interferometric measurements by Green (1994) are dominated by small-scale 
fluctuations, while filaments are features of the large-scale pattern. 
We also suspect, that mapping from
the line of observations to the velocity space of HI mitigates the anisotropy 
imposed by the regular magnetic field. Indeed, such a mapping inevitably 
entails averaging over the variations of
magnetic field direction along the line of sight. 

To understand the last point, one should recall, that one attempts 
to recover the information  using a galactic
rotation curve (see Kerr et al. 1986). The precision of this curve is
limited, apart from other factors, by the gas turbulent motion. Therefore
patches of gas at different distances from the observer
are mapped onto one point in the velocity space. If these patches
have different direction of magnetic field, the anisotropies of gas
density distribution are partially averaged out.

The mapping above is also important from the following reason. Verschuur
(1991a,b) studied his filaments in the data cube with one dimension
given by gaseous velocity. Arbitrary scaling along the $v$ axis in the
data cube can make initial round entities look elongated
or flat in the galactic space. In this paper, we are dealing with filaments
in the actual galactic space. Using galactic rotation curve it is possible
to find $v$ scaling in which a unit along $v$-axis will correspond
``on average'' to the same value in parsecs that a unit along $x$ or
$y$ direction in the sky plane. In this $xyv$ space the field will
be Gaussian and will have the same spectral index $\alpha$. Therefore
we expect to observe filaments in $xyv$ space, 
which, in fact, is done by Verschuur.
However, the {\it actual} filaments in $xyz$ space, where $z$ is the
axis align the line of observations may not coincide
with filaments in $xyv$ space.\footnote{This is a particular
case of the space-velocity mapping problem discussed in Burton (1992).}
All in all, we predict the existence of filaments that can be visualized
using contemporary data cubes, but warn, that the observed pattern is
distorted from the original one due to the problems intrinsic
to the velocity-space mapping (see Burton 1992).

Note, that the filaments discussed above are not related to any intermittent
turbulence, but appear in a purely Gaussian random field.
Surely, the effects of possible deviations from the Gaussian assumptions,
as well as dynamical formation of filaments, can be important in particular
regions of HI. Therefore our aim in this paper was not to explain all the 
the filaments observed in HI, but to attract the attention of the 
researchers in the field to a possibility of purely statistical origin
of filaments.

\section{Conclusions}

Our application of the theory of percolation of the Gaussian density
fields developed in Bardeen et al. (1986) and BKP96 resulted in predicting 
a purely statistical mechanism of filament formation in HI. For the 
shallow spectrum of  HI
density fluctuations obtained in L95 we found a pronounced large-scale 
 filamentary pattern that should arise in the galactic
atomic hydrogen due to this mechanism. The latter result corresponds to the
observations in V91a,b.

\noindent{\bf Aknowledgments}

A.L. acknowledges the support of  NASA grant
NAG5 2858. Authors are grateful to J. Bond and P. Martin for elucidating 
discussions. A.L. acknowledges very valuable comment by B.Draine.

\pagebreak
\begin{center}
{\bf Figure Captions}
\end{center}
{\bf Figure 1}. 
%N-body simulations with ... particles in the box with different regions
%at different overdensity contrast defined.  
A piece of random Gaussian field with the spectral index $-1$ 
and the different thresholds (actual overdensities). The a) figure depicts
isolated clusters of enhanced density at the density level $2.5 \sigma$.
As this ratio decreases, the clusters show connection by filaments.
For $2\sigma$ (b) the connections is established only 
between a few closest clusters but the number of connections increases
as $\sigma$ decreases (see (c) for the threshold equal to $1.5 \sigma$).
For the threshold   $1 \sigma$ shown in Fig.~1(d)  the random field
percolates and a 
network of filaments is easily observed. The field is generated with
periodic boundary conditions in a large box, of which the length of
the side of the box shown consitutes 0.625.\\

{\bf Figure 2}.
A bridge between two peaks with aligned ellipticities as a function of
distance. The height of the clusters is $\nu=1.7$, $e=0.32$ and $p=0$.
Contour is on the level $1\sigma$. The distance between peaks are 
(a) 3 times, (b) 4 times and (c) 5 times the peak scale (from top to
bottom). {\it Note}: scale of peak is defined as a filtering length $R_{pk}$
 with
"top-hat" filter (rather than gaussian, top-hat more closely
describes appearance of peak, top-hat with filter scale $R$
corresponds to average density in sphere of radius $R$). 
Crude rule is that top-hat with twice larger $R$ gives close result
as gaussian $R_{top-hat} = 2.2 R_{gaussian}$.\\

{\bf Figure 3}. The Gaussian field (left) and its reconstruction
with 20 largest peaks (right). The threshold shown at $2\sigma$ (a)
$1.5\sigma$ (b) and $1\sigma$ (c).\\

{\bf Figure 4}.
Formation of membrane-like structure for the Gaussian field with the
spectral index $-1$. The thresholds are the same as in Fig.1,
but in this region several peaks happen to occur nearby creating
the pattern shown. The length of the box side shown in the figure
is 0.8 of the box side of Fig.~1.
The field is generated with
periodic boundary conditions in a large box, of which the length of
the side of the box shown consitutes 0.5.\\

{\bf Figure 5}.
The patterns for different spectra of random density and for the
threshold $1\sigma$. The
figure (a) corresponds to the spectral index $0$ and shows
 cloud-like structures. More filamentary structures are revealed
by figure (b) corresponding to the spectral index $-1$.
Membranes are more vivid in  the figure (c) corresponding 
to the spectral index $-2$. The smoothing is the same as in Fig.~1.
and the spectra are normalized so that the dispersion $\sigma$ is
the same for all three cases. The random realization of the
amplitudes $a_k$ is chosen the same to visualize how the
structures are affected by the redistribution of power in the
spectrum. The whole box of simulation is shown.


\begin{thebibliography}{}
\bibitem{} Bally, J. 1996, Nature, 382, 114
\bibitem{} Bardeen, J.M., Bond, J.R., Kaiser, N., \& Szalay, A.S. 1986,
ApJ, 304, 15
\bibitem{} Bond, J.R. 1987a, Nearly Normal Galaxies From the Planck
Era to the Present, S.~Faber, Ner York: Springer-Verlag, 388
\bibitem{} Bond, J.R. 1987b, Cosmology \& Particle Physics, Workshop
Proc. Berkeley CA 1986, ed. I. Hidchcliffe, Singapore: World Scientific, 22
\bibitem{} Bond, J.R. 1989, Large-Scale Motions in the Universe, A Vatican
Study Week, V. Rubin \& G. Coyne, Princeton: Princeton University Press, 419
\bibitem{} Bond, J.R. \& Myers, S. 1995, a,b,c ApJS, (in press)
\bibitem{} Bond, J.R. \& Myers, S. 1995,d  CITA preprint
\bibitem{} Bond, J.R., Kofman, L. \& Pogosyan, D., 1996a, Nature, 380, 603
\bibitem{} Bond, J.R., Kofman, L.R.  \& Pogosyan, D.Yu., 1996b, CITA preprint
\bibitem{} Boss, A.P.  1987,  Phases of the Interstellar
Medium, in Interstellar Processes, eds Hollenbach D.J. and Thronson~H.A., 
 Reidel, Dordrecht, p.~321	
\bibitem{}  Burton, W.B., 1992,  Distribution and Observational 
Properties of the ISM, in: Pfenninger~D., Bartholdi~P. (eds.) 
\bibitem{} Carlquist, P. 1988, Ap\&SS, 144, 73
\bibitem{} Crovisier, J. \& Dickey, M. 1983, A\&A, 122, 282 
\bibitem{} Dickman, R.L.,  1985,  Turbulence in Molecular Clouds, 
in  Protostars and Planets II, eds Black~D.C. and Mathews~M.S., 
Tucson:  University of Arizona, 150
\bibitem{}   Dickman, R.L. \& Kleiner, S.C. 1985, ApJ, 295, 479
\bibitem{}  Elmegreen, B.G. 1992, 
Large Scale Dynamics of the Interstellar Medium, in 
The Galactic Interstellar Medium,  eds  
Pfenninger~D. and Bartholdi~P., Springer-Verlag, 157
\bibitem{}  Elmegreen, B.G. 1994, Infrared Cirrus and Interstellar
Diffuse Coulds, ed. Roc Cutri and Bill Latter, A.S.P., V58, 380
\bibitem{} Gammie, C.F. \& Ostriker, E.C. 1996, ApJ, 466 ... 
\bibitem{}  Gautier, T.N., Boulanger, 
F., P\'{e}rault, M. \& Puget, J.-L. 1992, AJ, {\bf 103}, 1313
\bibitem{} Gough, D.O., 1984, Phil. Trans. R. Soc. London, 
Ser.~A., {\bf 313}, 27
\bibitem{} Green, D.A. 1989, A\&A, 78, 277
\bibitem{} Green, D.A. 1993, MNRAS, 262, 328
\bibitem{} Green, D.A. 1994 Ap\&SS, 216, 61
\bibitem{} Heiles, C. in Polarimetry of the Interstellar Medium, 
eds Roberge W.G. and Whittet, D.C.B., 457
\bibitem{} Kalberla, P.M.W. \& Mebold, U. 1983, Mitt Astr. Ges., 58, 101
\bibitem{} Kalberla, P.M.W. \& Stenholm, L.G. 1983, 
Mitt. Astron. Ges., 60, 397 
\bibitem{} Kamp\'{e} de F\'{e}riet, J. 1955, 
in: Gas Dynamics of Cosmic Clouds, Amsterdam: North-Holland, 134 
\bibitem{} Kaplan, S.A. 1966, Interstellar Gas Dynamics, Pergamon, Oxford  
\bibitem{}  Kaplan, S.A. \& Pickelner, S.B. 1970, 
 The Interstellar Medium, Harvard University Press 
\bibitem{}  Kerr, F.J. \& Lynden-Bell, D. 1986, MNRAS, 221, 1023
\bibitem{} Kofman, L., Pogosyan, D., Shandarin, S. \& Melott, A. 1992, ApJ,
393, 437 
\bibitem{} Kolmogorov, A. 1941, Compt. Rend. Acad. Sci. USSR, 30, 301  
\bibitem{}  Kraichnan, R.H. 1965, Phys. Fluids, 8, 1385 
\bibitem{} Lazarian, A. 1992, Astron. and Astrophys. Transactions, 3, 33
\bibitem{} Lazarian, A. 1993a, Applied Scientific Research, 51, 191
\bibitem{} Lazarian, A. 1993b, Ap\&SS, 206, 37
\bibitem{} Lazarian, A. 1994, Plasma Phys. and Contr. Fusion, 36, 1013
 \bibitem{} Lazarian, A. 1995, A\&A, 293, 507
\bibitem{}  McKee, C.F. \&  Ostriker, J.P. 1977, ApJ, 218, 148
\bibitem{}  McCray, R. \& Snow, T.P., Jr. 1979, ARA\&A, 17, 213
\bibitem{} Monin, A.S. \& Yaglom, A.M. 1975, 
 Statistical Fluid Mechanics: Mechanics of Turbulence, vol. 2, The MIT Press 
\bibitem{} Mouschovias, T. Ch. 1991, ApJ, 373, 169
\bibitem{} Peebles, P.J.E. 1980, Large Scale Structure of the Universe,
Princeton: Princeton Univ. Press
\bibitem{} Rickett, B.J. 1991, ARA\& A, 28, 561
\bibitem{} Scalo, J.M. 1985, Fragmentation and Hierarchical 
Structure in the Interstellar Medium, in Protostars and Planets II, 
eds Black~D.C. and Mathews~M.S.,  Tucson: University of Arizona, 201
\bibitem{} Shandarin, S., 1983, Sov. Astr. Lett., 9, 104  
\bibitem{} Shull, J.M. 1987,  Phases of the Interstellar 
Medium, in Interstellar Processes, eds Hollenbach~D.J. and Thronson~H.A., 
 Reidel, Dordrecht, 225
\bibitem{} Spoelstra, T.A.Th. \& Brouw, W.A. 1976, A\&AS, 26, No.1	  
\bibitem{} Van Langevelde, H.J., Frail, D.A., Cordes, J.M. \& 
Diamond, P.J. 1992, ApJ, 396, 686
\bibitem{} Verschuur, G.L. 1991a, Ap\&SS, 185, 137
\bibitem{} Verschuur, G.L. 1991b, Ap\&SS, 185, 305
\bibitem{} Verschuur, G.L. 1995, Ap\&SS,
\bibitem{} Wiserman, J.J. \& Ho, P.T.P. 1996, Nature, 382, 139
\bibitem{} Zweibel, E.G. 1996, in Polarimetry of the Interstellar Medium, 
eds Roberge W.G. and Whittet, D.C.B., 486

\end{thebibliography}
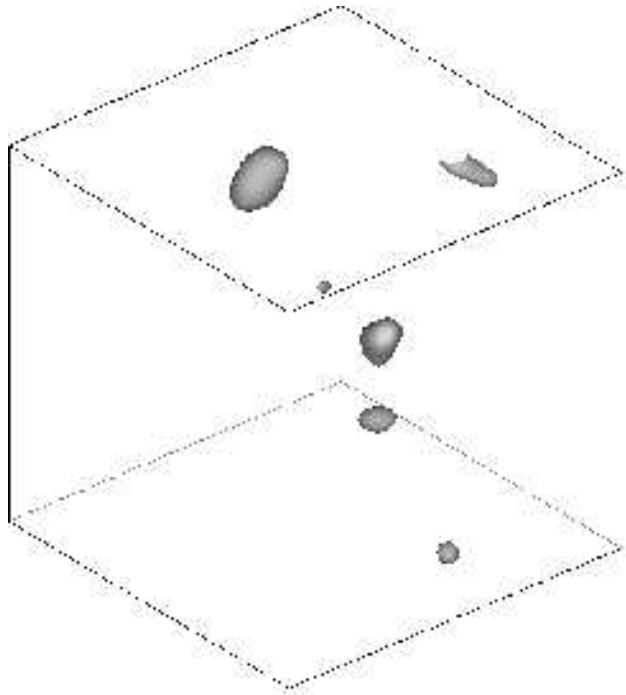
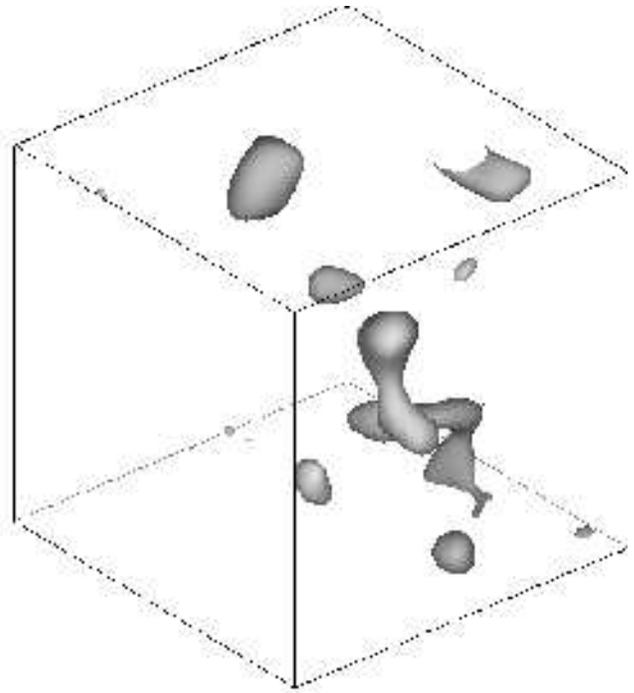
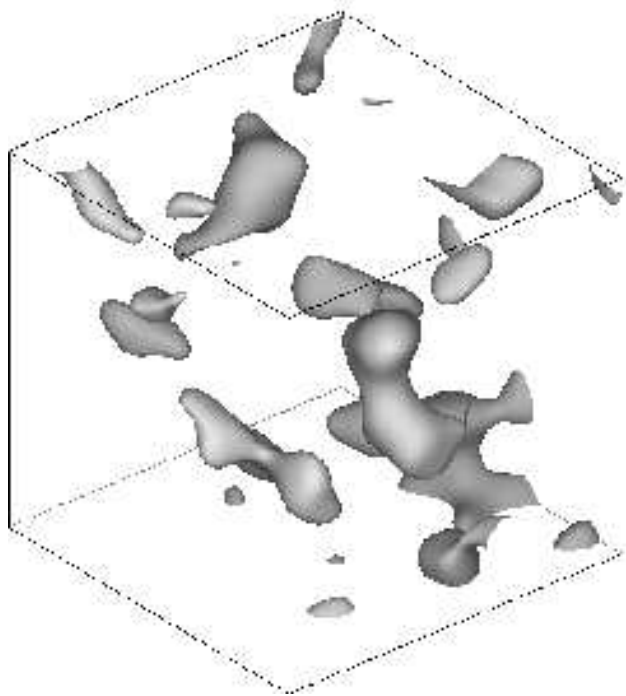
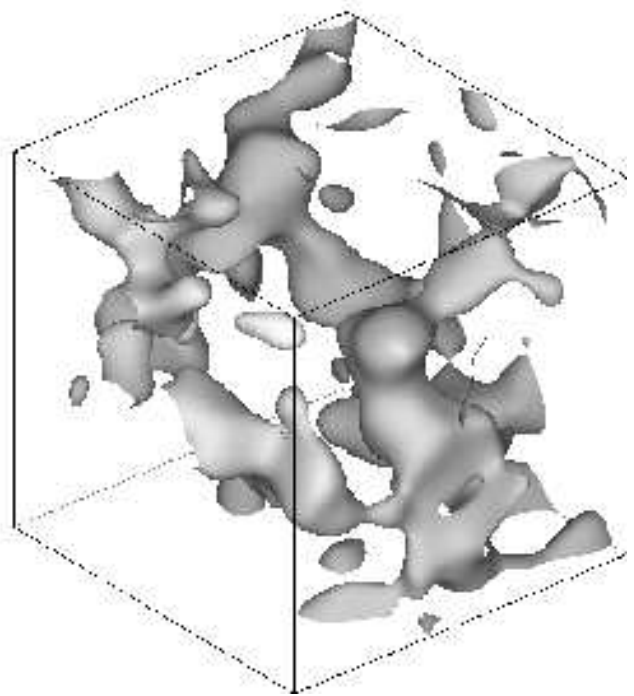
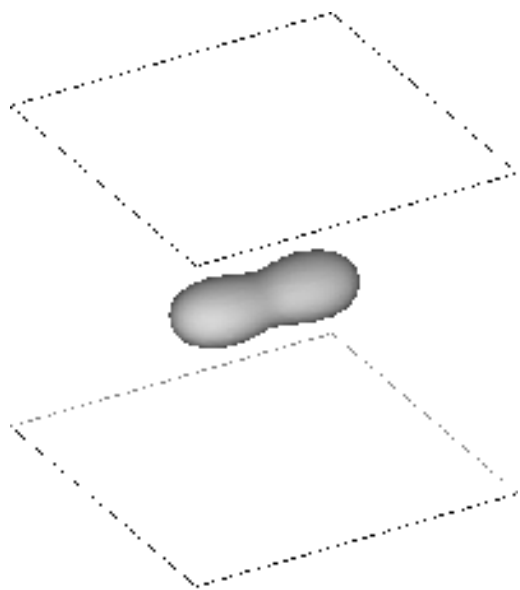
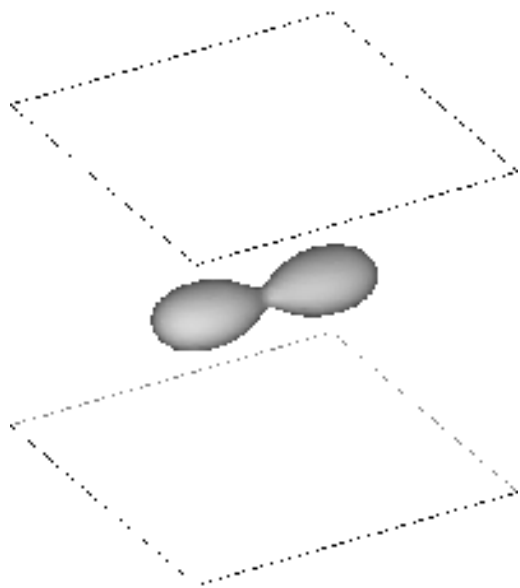
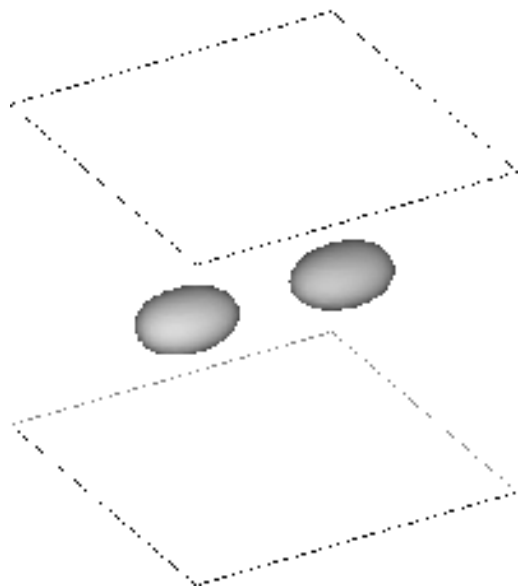
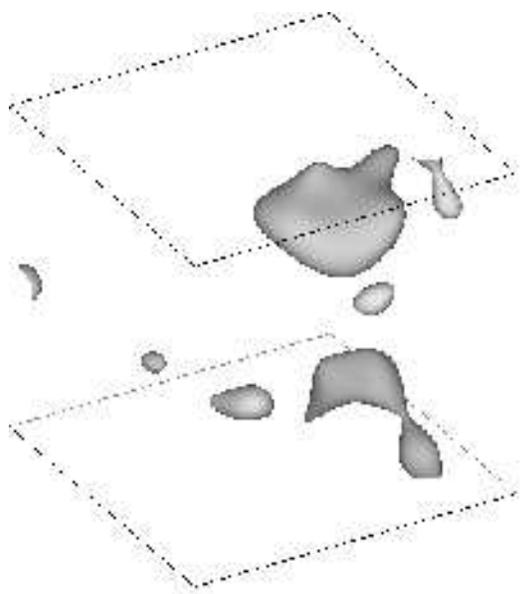
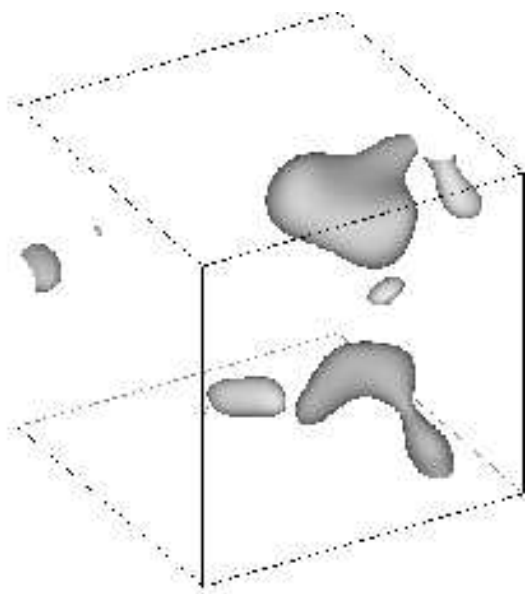
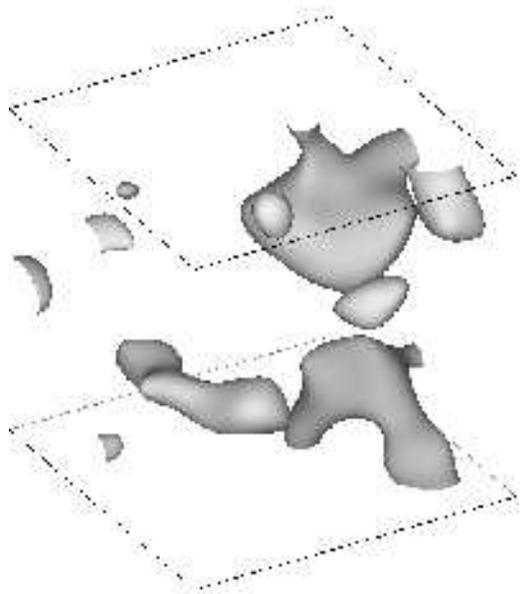
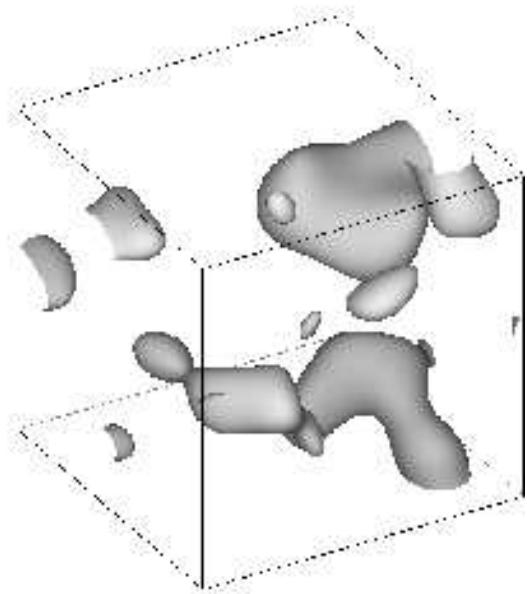
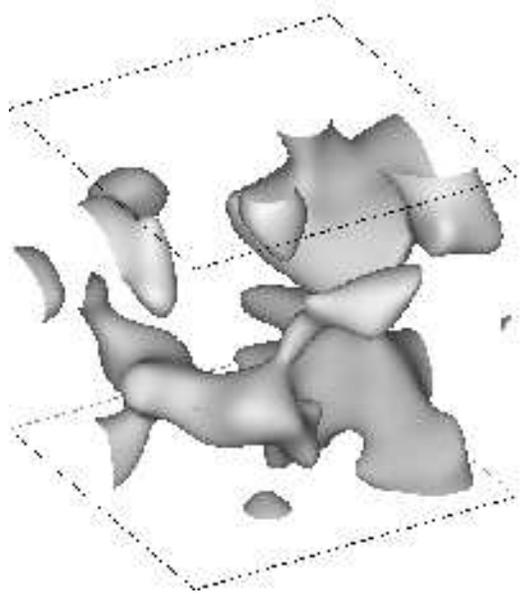
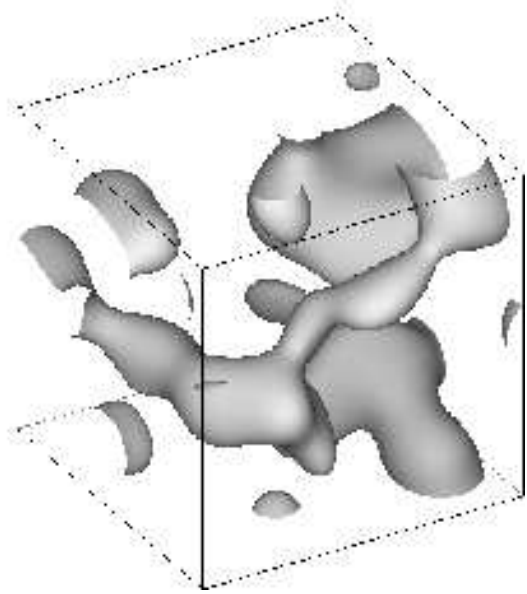
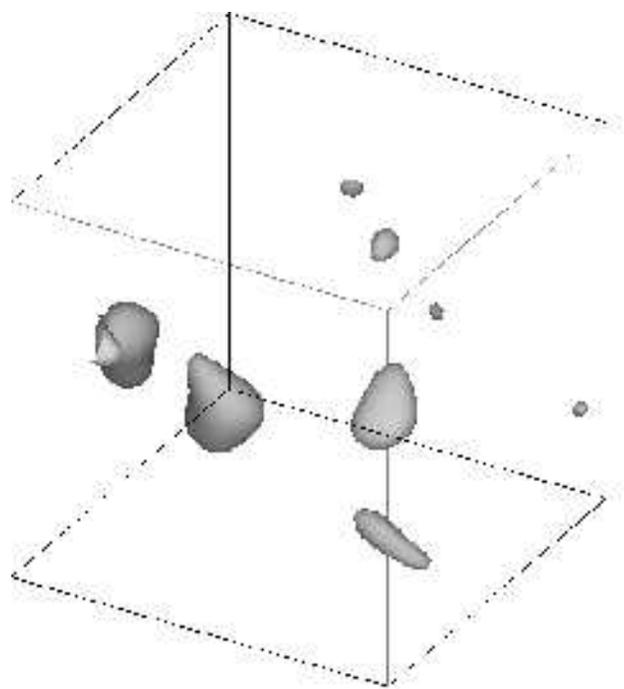
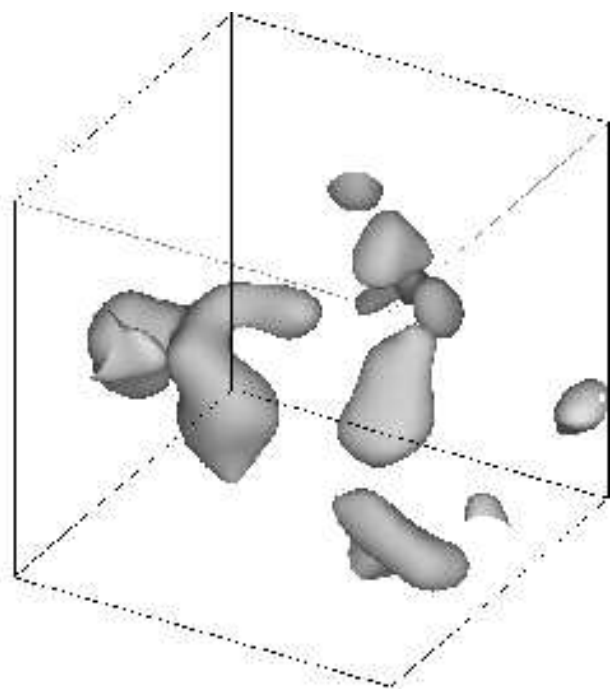
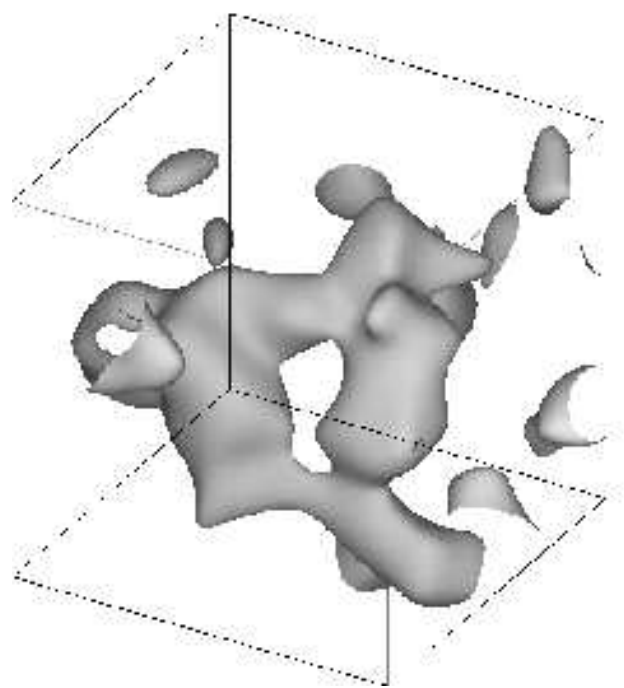
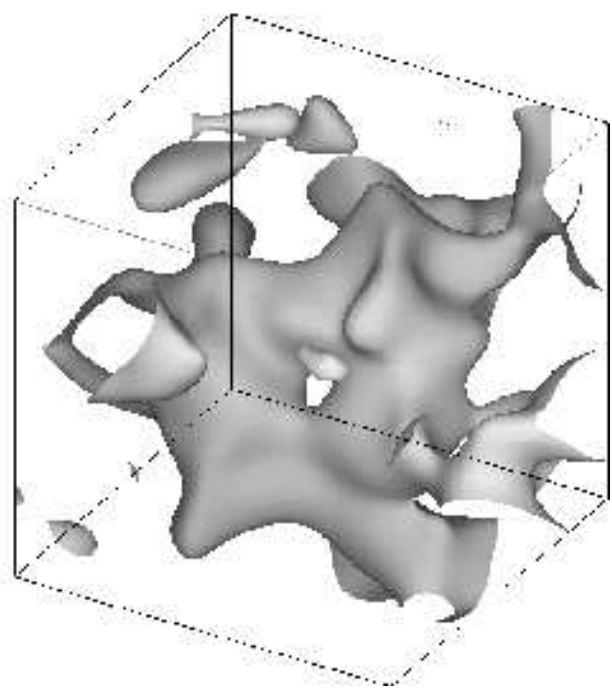
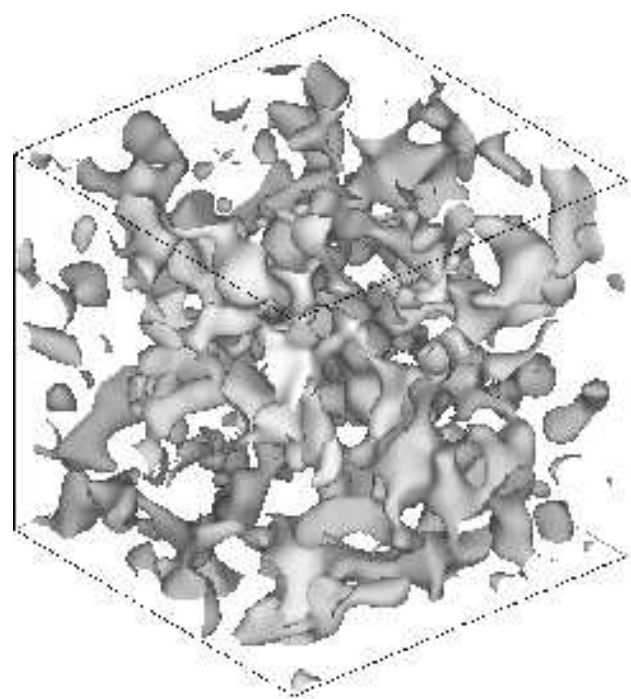
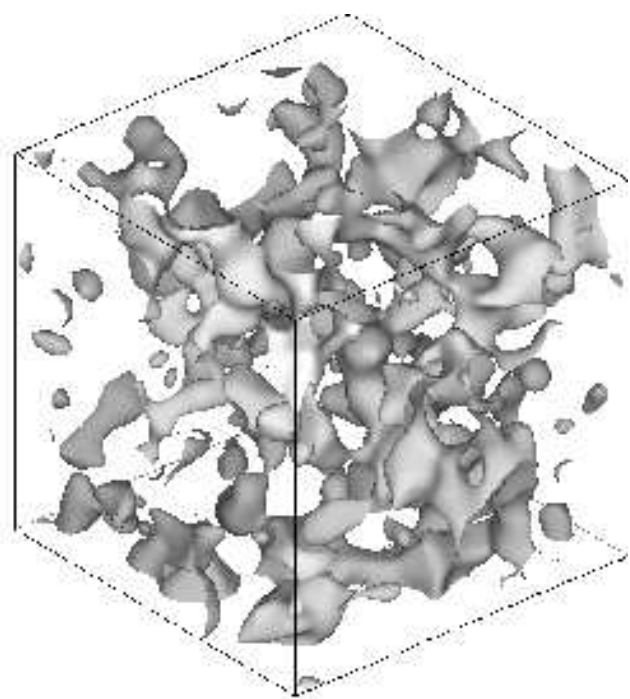
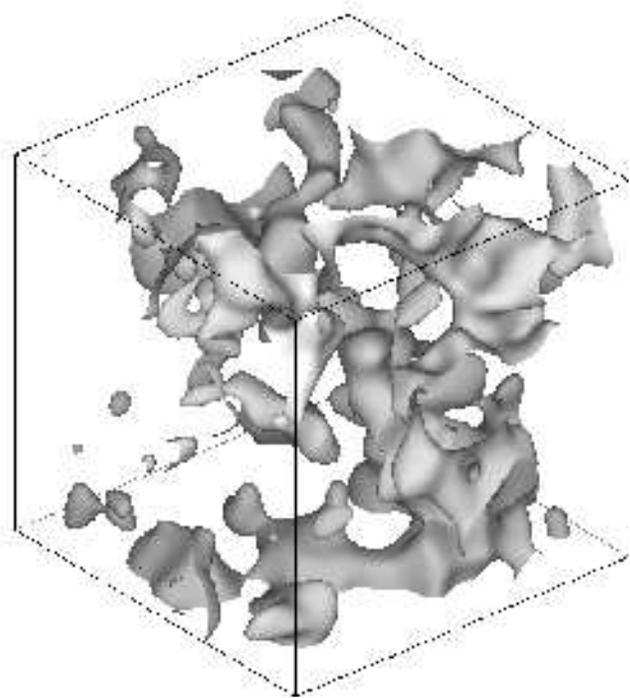
\end{document}